\documentclass[prd,amsmath,groupedaddress,twocolumn,eqsecnum,nofootinbib]{revtex4-2}
\usepackage{lipsum}
\usepackage{enumerate}
\usepackage{graphicx}
\usepackage{dcolumn}
\usepackage{subfig}
\usepackage{bm}
\usepackage{float}
\usepackage{hyperref}
\hypersetup{colorlinks=true, linkcolor=blue, citecolor=blue, urlcolor=blue}
\usepackage{caption}
\usepackage[hmargin={0.6in,0.3in},height=9.71in, top=0.7in
]{geometry}

\begin{document}

\title{ Gravitational Lensing Phenomena of Ellis-Bronnikov-Morris-Thorne Wormhole with Global Monopole and Cosmic String}

\author{Faizuddin Ahmed}
\email{faizuddinahmed15@gmail.com}
\affiliation{Department of Physics, The Assam Royal Global University, Guwahati, 781035, Assam, India
}

\author{\.{I}zzet Sakall{\i}}
\email{izzet.sakalli@emu.edu.tr}
\affiliation{Department of Physics, Eastern Mediterranean University, Famagusta Northern Cyprus, 99628,  via Mersin 10, Turkiye
}
\author{Ahmad Al-Badawi}
\email{ahmadbadawi@ahu.edu.jo}
\affiliation{Department of Physics, Al-Hussein Bin Talal University, Ma'an, 71111, Jordan
}

\date{\today}

\begin{abstract}
In this paper, we theoretically investigate gravitational lensing within the space-time framework of traversable wormholes, focusing on the combined effects of a global monopole  and a cosmic string. Specifically, we examine the Ellis-Bronnikov-Morris-Thorne wormhole metric and analyze how these topological defects influence photon trajectories. By considering the weak-field limit, we derive analytical expressions for the photon deflection angle, highlighting how factors such as the wormhole throat radius, the global monopole charge, and the cosmic string influence the gravitational lensing phenomenon. We also examine the weak-field limit of lensing phenomena for a zero wormhole throat radius and derive an analytical expression for the deflection angle of photon light in this scenario.
\end{abstract}



\maketitle

{\small

\section{Introduction}\label{introduction}

Gravitational lensing, a phenomenon predicted by Einstein's general theory of relativity \cite{isz01,isz02,isz03,isz03v1}, occurs when the trajectory of light from a distant source is bent by the gravitational field of a massive object situated between the source and the observer. This bending of light can lead to intriguing visual effects, such as the formation of multiple images of the source, Einstein rings, and arcs \cite{isz04,isz05}. Gravitational lensing has become a powerful tool in astrophysics and cosmology, providing valuable information about the distribution of matter in the universe, the properties of dark matter, and the expansion history of the cosmos \cite{isz06,isz07,isz08,isz08v1}. Recent studies on gravitational lensing around black holes have provided valuable insights into their structure and the behavior of space-time near them (see, for examples, \cite{fa2,fa3,fa4,fa5,fa6}). One of the most fascinating predictions of general relativity is the existence of wormholes, hypothetical tunnels that connect different regions of space-time or even different universes \cite{isz09,isz10}. Traversable wormholes, those that would allow passage through these space-time tunnels, require exotic matter with negative energy density to prevent their collapse \cite{MT1,MT2}. Among the various wormhole models, the Ellis-Bronnikov-Morris-Thorne wormhole (EBMT) stands out due to its simplicity and potential for traversability \cite{isz16,isz17}. Recent research on gravitational lensing associated with wormholes, referenced in \cite{fa8, fa9, fa10}, has unveiled novel understandings regarding the unique optical phenomena produced by these unusual formations. This study focuses on the static Ellis wormhole, a special case of the EBMT solution, characterized by the absence of an ether flow and, consequently, gravity within the wormhole itself. 

The general form of the Morris-Thorne wormhole metric is given by \cite{MT1,MT2}:
\begin{equation} \label{a1}
ds^2=-e^{2\,\Phi(r)}\,dt^2+\Big(1-\frac{A(r)}{r}\Big)^{-1}\,dr^2+r^2\,(d\theta^2+\sin^2 \theta\,d\phi^2),
\end{equation}
where $\Phi(r)$ is the redshift function and $A(r)$ is the shape function that determine structure of the wormhole. The range of the coordinates are $-\infty < t < +\infty$, $r \in [r_\text{min},\,\infty)$, $0 \leq \theta \leq \pi$, and $0 \leq \phi \leq 2\,\pi$, with $r=r_\text{min}$ is the radius of the wormhole throat. The EBMT-type wormhole space-time emerges when we set $\Phi'(r)=0$ (no redshift) and choose the shape function $A(r)=b^2/r$, where $b$ is a constant representing the throat radius of the wormhole \cite{MT1,MT2,JG}. This gives us the metric:
\begin{equation}\label{a2}
ds^2=-dt^2+\Big(1-b^2/r^2\Big)^{-1}\,dr^2+r^2\,(d\theta^2+\sin^2 \theta\,d\phi^2).
\end{equation}

By transforming $r=\sqrt{x^2+b^2}$, we can rewrite metric (\ref{a2}) as:
\begin{equation}\label{a3}
ds^2=-dt^2+dx^2+(x^2+b^2)\,(d\theta^2+\sin^2 \theta\,d\phi^2.
\end{equation}
The gravitational lensing phenomena in this wormhole geometry (\ref{a3}) has been studied in Refs. \cite{VP,KKN}. 

In Ref. \cite{JRN}, the authors studied nonlinear $\sigma$-models minimally coupled to Eddington-inspired Born-Infeld (EiBI) gravity. The action of the EiBI theory can be written as
\begin{equation}
    S_\text{EiBI}=\frac{1}{8\,\pi\,G\,\epsilon}\,\int\,d^4x\,\left[\sqrt{-|g_{\mu\nu}+\epsilon\,R_{\mu\nu}|}-\sigma\,\sqrt{-|g_{\mu\nu}|}\right]+S_{M}[g_{\mu\nu},\Phi].\label{action1}
\end{equation}
Here $G$ is Newton’s gravitational constant, $\epsilon$ is a parameter with dimension of area that controls the nonlinearity of the theory, called Eddington parameter. The term $S_{M}[g_{\mu\nu},\Phi]$ represents the action of the matter fields $\Phi$. In general, the constant $\sigma$ defines an effective cosmological constant, $\sigma=1+\epsilon\,\Lambda$. Throughout the analysis, $\Lambda=0$, and hence, $\sigma=1$. The action of the matter fields $\Phi$ is given by
\begin{eqnarray}
    S_M=\int\,\sqrt{-g}\,d^4x\,\left[-\frac{1}{2}\,(\partial_{\mu}\Phi^i)(\partial^{\mu}\Phi^i)-\frac{\lambda}{4}\,(\Phi^{i}\,\Phi^{i}-\eta^2)^2\right],\label{action2}
\end{eqnarray}
where $\Phi^i$ corresponds to a triplet of coupled real scalar fields. The model (\ref{action2}) displays spontaneous symmetry breaking $O(3) \to U(1)$. Here, the constants $\lambda$ and $\eta$ are, respectively, the dimensionless coupling constant and the energy scale of the spontaneous symmetry breaking.

The space-time describing a static and spherically symmetric solution is given by (please see Eq. (36) in Case A, section IV in Ref. \cite{JRN}, where $\kappa^2=8\,\pi\,G$)
\begin{eqnarray}
    ds^2&=&-\left(1-\kappa^2\,\eta^2-\frac{2\,M}{\sqrt{r^2+\epsilon\,\kappa^2\,\eta^2}}\right)\,dt^2\nonumber\\
    &+&\frac{r^2}{r^2+\epsilon\,\kappa^2\,\eta^2}\,\left(1-\kappa^2\,\eta^2-\frac{2\,M}{\sqrt{r^2+\epsilon\,\kappa^2\,\eta^2}}\right)^{-1}\,dr^2\nonumber\\
    &+&r^2\,(d\theta^2+\sin^2 \theta\,d\phi^2).\label{action3}
\end{eqnarray}
Noted that for $\epsilon>0$, the above solution (\ref{action3}) was first reported in Ref. \cite{RDL} (see Equations (30)--(32) there). 

However, in the context of wormhole geometry, the parameter $\epsilon<0$, that is $\epsilon=-|\epsilon|$, and therefore, the minimum value attain by $r$ is given as $r_\text{min}=\sqrt{|\epsilon|}\,\kappa\,\eta$, where $r_\text{min}$ is the wormhole throat radius. Setting the constant $M=0$ (see the appendix in Ref. \cite{JRN}), from Eq. (\ref{action3}) one can obtain
\begin{eqnarray}
    ds^2=-\left(1-\kappa^2\eta^2\right)dt^2+\frac{r^2\left(1-\kappa^2\eta^2\right)^{-1}}{(r^2-|\epsilon|\kappa^2\eta^2)}dr^2+r^2(d\theta^2+\sin^2 \theta d\phi^2),\quad\quad\label{action4}
\end{eqnarray}
which can be rewritten after scaling the temporal coordinate $t$ as follows (setting $\alpha^2=1-\kappa^2\,\eta^2=1-8\,\pi\,G\,\eta^2)$):
\begin{eqnarray}
    ds^2=-dt^2+\frac{dr^2}{\alpha^2\,\left(1-\frac{b^2}{r^2}\right)}+r^2\,(d\theta^2+\sin^2 \theta\,d\phi^2). \label{action5}
\end{eqnarray}
Here $\alpha$ is the global monopole (GM) parameter which characterizes the strength of the topological defect, with $\alpha < 1$ corresponding to the presence of a GM charge in the space-time. Also, it is worth noting that a massive monopole metric ($\epsilon>0$) was given in Eq. (35) in Ref. \cite{RDL} as follows:
\begin{eqnarray}
    ds^2=-dt^2+\frac{dr^2}{\left(1+\frac{\epsilon}{r^2}\right)}+\alpha^2\,r^2\,(d\theta^2+\sin^2 \theta\,d\phi^2).\label{action6}
\end{eqnarray}

The potential coexistence of wormhole, GM, and cosmic string (CS) within the universe serves as the foundation for this research, suggesting that their combined gravitational phenomena may produce distinct observational characteristics. This paper explores the interaction between a GM and CS, specifically focusing on their influence on the trajectory of light as it traverses the space-time geometry associated with a static Ellis wormhole. GM \cite{MBAV} are hypothetical point-like defects formed due to the spontaneous breaking of a global symmetry. They possess a non-trivial topology and can exert significant gravitational influence, causing light rays to deflect. CS, on the other hand, are hypothetical one-dimensional topological defects that may have formed during symmetry-breaking phase transitions in the early universe \cite{isz21,isz22,isz26}. These strings are incredibly dense and possess high tension, leading to significant gravitational effects, including the deflection of light \cite{isz28,isz29}. By investigating the deflection of light in this scenario, we aim to explore how the presence of a GM and CSs might alter the wormhole's space-time geometry and affect the trajectories of light rays and to show that there is a potential for detecting these objects through their combined lensing effects. This involves analyzing the distinctive lensing patterns that might arise from the interplay of these objects and assessing the feasibility of observing these patterns with current or future telescopes. To achieve these objectives, we employ a combination of analytical calculations and numerical simulations. We start by deriving the equations of motion for photons in the combined gravitational field of the wormhole, GM, and CS. These equations are then solved analytically to determine the trajectories of light rays and calculate the deflection angle. The numerical simulations will take into account the parameters of the wormhole (throat radius), the GM (deficit solid angle), and the CS (tension and distribution).

The paper is organized as follows: In Section \ref{sec2}, we detail the theoretical framework for calculating the deflection angle of light in the presence of a GM and CS in a static Ellis wormhole space-time. Then, we apply this formalism to specific cases and analyze how the deflection angle depends on the parameters of the wormhole, the GM, and the CS. In Section \ref{sec3}, we discuss the observational implications of our findings and the potential for detecting these objects through their lensing signatures. Finally, in Section \ref{sec4}, we summarize our results and conclude with a discussion of future research directions. 

\section{Deflection of Photon Light in EBMT-type Wormhole with GM and CS}\label{sec2}

In this article, we explore the deflection of photon light in the background of EBMT-type wormholes, incorporating GM and CS. We investigate how various parameters influencing the wormhole geometry affect the photon deflection angle. Specifically, we examine two distinct forms of the EBMT-type wormhole metric, which account for these topological defects, and analyze their impact on the deflection phenomenon.

\begin{center}
    {\bf Metric-I}
\end{center}

Inspired by the topologically charged wormhole metric (\ref{action5}) and the Aryal-Ford-Vilenkin solution \cite{AFV}, which describes a CS passing through a black hole, we present a topologically charged traversable wormhole space-time pierced by a CS given by
\begin{eqnarray}
   ds^2=-dt^2+\frac{dr^2}{\alpha^2\,\left(1-\frac{b^2}{r^2}\right)}+r^2\,(d\theta^2+\beta^2\,\sin^2 \theta\,d\phi^2),\label{b1}
\end{eqnarray}
where $\beta=(1-4\,G\,\mu)$ is a CS parameter with $\mu$ being mass per unit length of the strings \cite{PSL1,PSL2,PSL3} and others are stated earlier. In the limit when $\beta=1$, one can recover a topologically charged wormhole metric \ref{action5}. Moreover, in the limit when $\alpha=1$, we recover a EBMT-type wormhole metric with a CS \cite{FA}. Although the metric is singular at $r=b$, the proper radial distance/length for the wormhole metric is given by $\ell(r)=\pm\,\frac{\sqrt{r^2-b^2}}{\alpha}=\frac{1}{\alpha}\,\ell_\text{EBMT}(r)$, a real quantity. This proper distance is equal to $\frac{1}{\alpha}$ times the proper radial distance obtained in EBMT wormhole. The most crucial point is that if we compare space-time (\ref{b1}) with the standard form of Morris-Thorne wormhole metric given in Eq. (\ref{a1}) with a CS, then one can find $\Phi'(r)=0$ and the shape function $A(r)=r\,\left[1-\alpha^2\,\left(1-\frac{b^2}{r^2}\right)\right]$. Because of GM charge, the wormhole metric (\ref{b1}) is not asymptotically flat globally, unlike the Ellis-Bronnikov-Morris-Throne-type wormhole metric (\ref{b2}) since $\lim_{r \to \infty}\,\frac{A(r)}{r} \to 1-\alpha^2$. However, asymptotically, $r \to \infty$, the metric (\ref{b1}) is locally flat conical space-time generated by a CS including a GM. Another point is that the above wormhole metric (\ref{b1}) is not spherically symmetric since the factor  $\beta^2\,\sin^2 \theta$ causes the $\phi$-direction to have a different scaling compared to the $\theta$-direction, which means the geometry in the $\phi$-direction is not the same as in the $\theta$-direction. In other words, the factor $\beta^2\,\sin^2\theta$ introduces a non-uniform scaling in the $\phi$-direction. This breaks the rotational symmetry of the space-time, and hence, the 2-dimensional surface is not isotropic, but anisotropic in nature. 

Consider the 2-dimensional surface of constant $t$ and constant $r$ (say unity), which we will refer to as the “unit defect sphere.” It is evident that in conical space, this surface is not spherically symmetric. Specifically, the circumference of a circle in the equatorial plane ($\theta =\pi/2$) is $2\,\pi\,\beta$, while the circumference of any meridianal circle ($\phi$=\text{const.}) is $2\,\pi$. The total area of this surface is given by $4\,\pi\,\beta$. For a realistic cosmic string, where $\beta < 1$, one can visualize the defect sphere as the standard sphere in Euclidean space with a segment of azimuthal length $2\,\pi (1 - \beta)$ removed, and the corresponding boundary points identified. Globally, the defect sphere is axially symmetric, but locally it possesses three Killing vectors \cite{DVG}
\begin{eqnarray}
    \vec{\xi}_1&=&\sin(\beta\phi)\,\hat{\partial}_{\theta}+\frac{1}{\beta}\,\cos (\beta \phi)\,\cot \theta\,\hat{\partial}_{\phi},\nonumber\\
    \vec{\xi}_2&=&-\cos (\beta\phi)\,\hat{\partial}_{\theta}+\frac{1}{\beta}\,\sin (\beta \phi)\,\cot \theta\,\hat{\partial}_{\phi},\nonumber\\
    \vec{\xi}_3&=&-\frac{1}{\beta}\,\hat{\partial}_{\phi}.\label{killing1}
\end{eqnarray}
These vectors forming an algebra $O(3)$ as
\begin{equation}
    [\vec{\xi}_i\,,\,\vec{\xi}_j]=\varepsilon_{ijk}\,{\vec \xi}_k\quad\quad\quad (i,j,k=1,2,3).\label{killing}
\end{equation}

Now, we study geometric motions of test particles and discuss the results. Moreover, we study deflection of photon light and show how various parameter involve in the wormhole geometry influence the bending of photon light. The Lagrangian density function is defined by
\begin{equation}
\mathcal{L}=\frac{1}{2}\,g_{\mu\nu}\,\left(\frac{dx^{\mu}}{d\tau}\right)\,\left(\frac{dx^{\nu}}{d\tau}\right),\label{b2}
\end{equation}
where $\tau$ is the affine parameter of the curve, and $g_{\mu\nu}$ is the metric tensor. 

For the space-time (\ref{b1}), the Lagrangian density function reads:
\begin{eqnarray}
\mathcal{L}&=&\frac{1}{2}\,\Bigg[-\left(\frac{dt}{d\tau}\right)^2+\alpha^{-2}\,\Big(1-b^2/r^2\Big)^{-1}\,\left(\frac{dr}{d\tau}\right)^2+r^2\,\left(\frac{d\theta}{d\tau}\right)^2\nonumber\\
&+&r^2\,\beta^2\,\sin^2\theta\,\left(\frac{d\phi}{d\tau}\right)^2\Bigg].\label{b3}
\end{eqnarray}

The constants of motion are the energy $\mathrm{E}=\xi^{\mu}_{(t)}\,p_{\mu}=-p_t$ and the component of the angular momentum of the test particle that is aligned with the axis of the string (here the $z$-axis) given by $L_z=\xi^{\mu}_{(\phi)}\,p_{\mu}=p_{\phi}/\beta$ \cite{DVG} and the square of the angular momentum $\mathrm{L}^2=p^2_{\theta}+p^2_{\phi}\,(\beta\,\sin \theta)^{-2}$. In our case, $p_{\theta}=r^2\,(d\theta/d\tau)$ and $p_{\phi}=\beta^2\,r^2\,\sin^2 \theta\,(d\phi/d\tau)$. Therefore, we find the following set of geodesic equation of motion
\begin{eqnarray}
    \left(\frac{dt}{d\tau}\right)^2&=&\mathrm{E}^2,\label{b4}\\
    \left(\frac{d\phi}{d\tau}\right)^2&=&\frac{\mathrm{L}^2_z}{\beta^2\,r^4\,\sin^4\theta},\label{b5}\\
    \left(\frac{d\theta}{d\tau}\right)^2&=&\frac{\mathrm{L}^2}{r^4}-\frac{\mathrm{L}^2_z}{r^4\,\sin^2 \theta}.\label{b5aa}
\end{eqnarray}

From the Lagrangian density function (\ref{b3}), we find
\begin{equation}
    \left(\frac{dr}{d\tau}\right)^2=\alpha^2\,\left(1-\frac{b^2}{r^2}\right)\,\left(\varepsilon+\mathrm{E}^2-\frac{\mathrm{L}^2}{r^2}\right),\label{b6}
\end{equation}
where $\varepsilon=0$ for light-like geodesics and $-1$ for time-like.

The above equation (\ref{b6}) can be rewritten as
\begin{equation}
    \alpha^{-2}\,\left(1-\frac{b^2}{r^2}\right)^{-1}\,\left(\frac{dr}{d\tau}\right)^2+V_\text{eff}(r)=\mathrm{E}^2,\label{b6aa}
\end{equation}
where the effective potential $V_\text{eff}(r)$ is given by
\begin{equation}
    V_\text{eff}(r)=-\varepsilon+\frac{\mathrm{L}^2}{r^2}.\label{b6bb}
\end{equation}

Moreover, Eq. (\ref{b5aa}) can be rewritten as
\begin{equation}
    r^4\,\left(\frac{d\theta}{d\tau}\right)^2+V_\text{eff}(\theta)=\mathrm{L}^2,\label{b6cc}
\end{equation}
where the effective potential $V_\text{eff}(\theta)$ is given by
\begin{equation}
    V_\text{eff}(\theta)=\frac{\mathrm{L}^2_z}{\sin^2 \theta}.\label{b6dd}
\end{equation}

Since $\mathrm{L}^2 \geq V_\text{eff}(\theta)$, the motion of test particle is restricted
\begin{equation}
    \arcsin(\mathrm{L}_z/\mathrm{L}) \leq \theta \leq \pi-\arcsin(\mathrm{L}_z/\mathrm{L}).\label{b6ee}
\end{equation}
Apparently, for $\mathrm{L}_z=\mathrm{L}$, the motion occurs in the equatorial plane, {\it i.e.}, $\theta=\pi/2$.

Now, excluding from Eqs. (\ref{b5}) and (\ref{b5aa}) the radial coordinate $r$, we obtain
\begin{equation}
    \frac{d\theta}{d\phi}=\beta\,\sin \theta\,(\zeta^2\,\sin^2 \theta-1)^{1/2},\label{b6ff}
\end{equation}
where $\zeta^2=\mathrm{L}^2/L^2_z$.  This is the geodesic equation on the unit defect sphere \cite{DVG}. Simplification of the Eq. (\ref{b6ff}) result
\begin{equation}
    \cot^2 \theta=(\zeta^2-1)\,\sin^2 (\beta\,\phi).\label{b6gg}
\end{equation}
If $\zeta=1 \Rightarrow \mathrm{L}_z=\mathrm{L}$, then from the above equation, we have $\cot^2 \theta=0\Rightarrow \theta=\pi/2$, and hence, the motion occurs in the equatorial plane.

Now, we focus on $r(\phi)$ motion, that is, change of $r$ with respect to the change of $\phi$. Using Eqs. (\ref{b5}) and (\ref{b6}), we find
\begin{eqnarray}
    \left(\frac{dr}{d\phi}\right)^2=\frac{\left(\varepsilon+\mathrm{E}^2-\mathrm{L}^2/r^2\right)}{\mathrm{L}^2_z}\,\frac{\alpha^2\,\beta^2\,r^4\,(1-b^2/r^2)}{[1+(\zeta^2-1)\,\sin^2 (\beta\,\phi)]^2}.\label{b6hh}
\end{eqnarray}
For orbits with $\phi \neq \mbox{const.}$, the presence of the deficit angle influences the shape of the orbits significantly. The orbits are in general nonplanar (except for the case $\mathrm{L}_z=\mathrm{L}$) and lie in a plane that has $\vec{\mathrm{L}}$ as its normal. Since $\vec{\mathrm{L}}$ is not conserved, this plane precesses which was explain in detailed in Ref.\cite{DVG}. 

From the equation (\ref{b6hh}) above, it is evident that solving this equation is quite challenging, making it difficult to determine the deflection angle of photon light. As discussed in Ref. \cite{EH} (specifically in Section III, Case A), to examine the effect of CS parameter $\beta$ on the deflection of light ($\varepsilon = 0$), we must set $\zeta=1$ which implies $\mathrm{L}_z= \mathrm{L}$. This condition result $\theta = \pi/2$, and hence, motion of photon trajectories take place on the equatorial plane. A similar argument is adopted in Refs. \cite{CL,SG}, where the authors investigated the gravitational lensing and/or bending of light phenomena on the equatorial plane in the background of a Schwarzschild black hole pierced by a CS and in a Schwarzschild black hole in the context of a string cloud.  

Thereby, considering photon trajectories on the equatorial plane, we can rewrite Eq. (\ref{b6hh}) as
\begin{eqnarray}
    \left(\frac{dr}{d\phi}\right)^2=\frac{\alpha^2\,\beta^2}{\gamma^2}\,(r^2-b^2)\,(r^2-\gamma^2),\label{b6ii}
\end{eqnarray}
where $\gamma=\mathrm{L}/\mathrm{E}$ is the impact parameter for photon light.

The above Eq. (\ref{b6ii}) can be rewritten as
\begin{equation}
    \frac{dr}{d\phi}=\frac{\alpha\,\beta}{\gamma}\,\sqrt{(r^2-b^2)\,(r^2-\gamma^2)}.\label{b10}
\end{equation}

Therefore, the angular deflection $\delta\,\phi$ is given by 
\begin{equation}
    \delta\phi=\Delta\phi-\pi,\label{b11}
\end{equation}
where
\begin{equation}
     \Delta\phi=\frac{2\,\gamma}{\alpha\,\beta}\,\int^{\infty}_{r_0=\gamma}\,\frac{dr}{\sqrt{(r^2-b^2)\,(r^2-c^2)}}.\label{b14}
\end{equation}
Here $r_0$ is the closest approximation radius that a photon coming from a source at infinity may turn and then escapes to a faraway observer. This radius can be obtained when $\dot{r}=0$ in Eq. (\ref{b6}) which implies that $r_0=\gamma$.

Let’s introduce a new variable $w=\frac{\gamma}{r}$ where, $dw=-\frac{dr}{r^2}$ and $g=b/\gamma$ in the above equation (\ref{b14}). Note that as $r \to \gamma$, $w \to 1$ and as $r \to \infty$, we have $w \to 0$. Therefore, in terms of $w$, from Eq. (\ref{b14}), we obtain
\begin{equation}
     \Delta\phi=\frac{2}{\alpha\,\beta}\,K(g),\quad\quad g=\frac{b}{\gamma},\label{b15}
\end{equation}
where
\begin{equation}
    K(g)=\int^{1}_{0}\,\frac{dw}{\sqrt{(1-w^2)\,(1-g^2\,w^2)}},\quad 0< g < 1\label{b16}
\end{equation}
is a complete elliptic integral of first kind \cite{MA}. In terms of the Gauss hypergeometric function, the complete elliptic integral of the first kind can be expressed as
\begin{equation}
    K(g)=\frac{\pi}{2}\,{}_2F_{1} \left(\frac{1}{2},\,\frac{1}{2},\,1,\,g^2\right).\label{relation}
\end{equation}

Defining $w=\sin \chi$ in the integral (\ref{b16}), we obtain the complete elliptic integral of the first kind by \cite{MA}
\begin{equation}
K(g)=\int^{\frac{\pi}{2}}_{0}\,\frac{d\chi}{\sqrt{1-g^2\,\sin^2 \chi}},\label{b17}
\end{equation}
where $g$ is called the modulus of the elliptic integral. 

In the weak field approximation, where the gravitational field is assumed to be weak, we consider the condition $g=b/\gamma<1$, which implies that $b <\gamma$. This condition suggests that the light ray does not pass through the wormhole throat. In other words, for the light to pass through the wormhole throat, the wormhole throat radius $b$ must be sufficiently large in comparison to the impact parameter $\gamma$. When $b$ is smaller than $\gamma$, the effective potential becomes prohibitive for the trajectory of the photon, preventing it from crossing the throat region. This restriction highlights the interplay between the topological features of the wormhole, the CS, and the geometry of the space-time in determining the path of light.

The expression of the complete elliptic integral is therefore, from Eq. (\ref{b17}) given by \cite{MA} 
\begin{eqnarray}
K(g)=\frac{\pi}{2}\sum^{\infty}_{n=0}\Bigg(\frac{(2n)!}{2^{2n}(n!)^2} \Bigg)^2g^{2n}
=\frac{\pi}{2}\Bigg[1+\frac{g^2}{4}+\frac{9g^4}{64}+\frac{25}{256}g^6+...\Bigg].\quad\quad\label{b18}
\end{eqnarray}

Hence, the deflection angle from (\ref{b11}) using (\ref{b15}), (\ref{relation}), and (\ref{b18}) is given by ($g=b/\gamma$)
\begin{eqnarray}
\delta\phi_{\text{Metric-I}}&=&\frac{\pi}{\alpha\,\beta}\,\,{}_2F_{1} \left(\frac{1}{2},\,\frac{1}{2},\,1,\,g^2\right)-\pi\nonumber\\
&=&\left(\frac{1}{\alpha\,\beta}-1\right)\,\pi+\frac{\pi}{4\,\alpha\,\beta}\,g^2+\frac{9\,\pi}{64\,\alpha\,\beta}\,g^4+\frac{1}{\alpha\,\beta}\,\mathcal{O} (g)^6.\quad\quad\label{b19}
\end{eqnarray}

One can see that the deflection angle of photon light depends on both the GM ($\alpha$) and CS parameters ($\beta$), and thus, gets modified.

\begin{center}
    {\bf Metric-II} 
\end{center}

Here, we present another form of an EBMT-type traversable wormhole in the presence of a GM and CS. The line-element describing a traversable wormhole space-time with these topological defects is given by
\begin{eqnarray}
   ds^2=-dt^2+\frac{dr^2}{\left(1-\frac{b^2}{r^2}\right)}+\alpha^2\,r^2\,(d\theta^2+\beta^2\,\sin^2 \theta\,d\phi^2),\label{f1}
\end{eqnarray}
where the symbols have their usual meanings as stated earlier. Here, the shape function form is $A(r)=\frac{b^2}{r}$, and thus, the wormhole metric (\ref{f1}) with a GM and CS is asymptotically flat: $\lim_{r \to \infty}\,\frac{A(r)}{r} \to 0$. In the limit when $b=0$ and $\alpha=1$, one can recover a CS space-time \cite{AV,Vilekan,BL}. Moreover, in the limit $b=0$ and $\beta=1$, we recover a point-like GM \cite{MBAV}.  

Analogue to previous analysis, globally the unit defect sphere metric (\ref{f1}) is also only axially symmetric, locally it possesses three Killing vectors
\begin{eqnarray}
    {\vec \eta}_1&=&\frac{\sin(\beta \phi)}{\alpha}\,\hat{\partial}_{\theta}+\frac{1}{\alpha\,\beta}\,\cos (\beta \phi)\,\cot \theta\,\hat{\partial}_{\phi},\nonumber\\
    {\vec \eta}_2&=&-\frac{\cos (\beta \phi)}{\alpha}\,\hat{\partial}_{\theta}+\frac{1}{\alpha\,\beta}\,\sin (\beta \phi)\,\cot \theta\,\hat{\partial}_{\phi},\nonumber\\
    {\vec \eta}_3&=&-\frac{1}{\alpha\,\beta}\,\hat{\partial}_{\phi}.\label{killing3}
\end{eqnarray}

Now, we focus into the geodesic motion and then study bending angle of photon ray. For the space-time (\ref{f1}), the Lagrangian density function reads:
\begin{eqnarray}
\mathcal{L}&=&\frac{1}{2}\,\Bigg[-\left(\frac{dt}{d\tau}\right)^2+\left(1-\frac{b^2}{r^2}\right)^{-1}\,\left(\frac{dr}{d\tau}\right)^2+\alpha^2\,r^2\,\left(\frac{d\theta}{d\tau}\right)^2\nonumber\\
&+&\alpha^2\,\beta^2\,r^2\,\sin^2\theta\,\left(\frac{d\phi}{d\tau}\right)^2\Bigg].\label{f2}
\end{eqnarray}

The constants of motion are the energy $\mathrm{E}=\xi^{\mu}_{(t)}\,p_{\mu}=-p_t$ and the component of the angular momentum of the test particle that is aligned with the axis of the string (here the $z$-axis) given by $L_z=\xi^{\mu}_{(\phi)}\,p_{\mu}=p_{\phi}/(\alpha\,\beta)$, where in our case $p_{\phi}=\alpha^2\,\beta^2\,r^2\,\sin^2 \theta\,(d\phi/d\tau)$. Therefore, we find the following set of geodesic equation:
\begin{eqnarray}
    \left(\frac{dt}{d\tau}\right)^2&=&\mathrm{E}^2,\label{f3}\\
    \left(\frac{d\phi}{d\tau}\right)^2&=&\frac{\mathrm{L}^2_z}{\alpha^2\,\beta^2\,r^4\,\sin^4\theta},\label{f4}\\
    \left(\frac{d\theta}{d\tau}\right)^2&=&\frac{\mathrm{L}^2}{\alpha^2\,r^4}-\frac{\mathrm{L}^2_z}{\alpha^2\,r^4\,\sin^2 \theta}.\label{f5}
\end{eqnarray}

From the Lagrangian density function (\ref{f2}), we find
\begin{equation}
    \left(\frac{dr}{d\tau}\right)^2=\left(1-\frac{b^2}{r^2}\right)\,\left(\varepsilon+\mathrm{E}^2-\frac{\mathrm{L}^2}{r^2}\right).\label{f6}
\end{equation}
The above equation (\ref{f6}) can be rewritten as
\begin{equation}
    \left(1-\frac{b^2}{r^2}\right)^{-1}\,\left(\frac{dr}{d\tau}\right)^2+V_\text{eff}(r)=\mathrm{E}^2,\label{f7}
\end{equation}
where the effective potential $V_\text{eff}(r)$ is the same as given in Eq. (\ref{b6bb}).

Moreover, Eq. (\ref{f5}) can be rewritten as
\begin{equation}
    \alpha^2\,r^4\,\left(\frac{d\theta}{d\tau}\right)^2+V_\text{eff}(\theta)=\mathrm{L}^2,\label{f8}
\end{equation}
where the effective potential $V_\text{eff}(\theta)$ is the same as given in Eq. (\ref{b6dd}).

Now, excluding from Eqs. (\ref{f4}) and (\ref{f5}) the radial coordinate $r$, we obtain the same equation as obtained in Eq. (\ref{b6ff}) whose solution is given in Eq. (\ref{b6gg}). Analogue to the previous analysis, here also we set $\theta=\pi/2$ to study the effect of the CS and the GM on the deflection of photon rays. Thereby, the $r(\phi)$ motion in the equatorial plane is given by
\begin{eqnarray}
    \left(\frac{dr}{d\phi}\right)^2=\frac{\alpha^2\,\beta^2}{\gamma^2}\,(r^2-b^2)\,(r^2-\gamma^2).\label{f9}
\end{eqnarray}

Following the previous technique, one can find the deflection angle of photon ray given by
\begin{eqnarray}
\delta\phi_{\text{Metric-II}}&=&\frac{\pi}{\alpha\,\beta}\,\,{}_2F_{1} \left(\frac{1}{2},\,\frac{1}{2},\,1,\,g^2\right)-\pi\nonumber\\
&=&\left(\frac{1}{\alpha\,\beta}-1\right)\,\pi+\frac{\pi}{4\,\alpha\,\beta}\,g^2+\frac{9\,\pi}{64\,\alpha\,\beta}\,g^4+\frac{1}{\alpha\,\beta}\,\mathcal{O} (g)^6\quad\quad\label{f10}
\end{eqnarray}
which is similar to the expression given in Eq. (\ref{b19}) for the Metric-I.

From the above analysis of the deflection angle for the photon ray, it is clear that Metrics I and II yield the same expression for the angular deflection of light, that is, $\delta\phi_{\text{Metric-I}}=\delta\phi_{\text{Metric-II}}$.

\begin{center}
\begin{figure}[ht!]
\begin{centering}
\subfloat[]{\centering{}\includegraphics[width=0.7\linewidth]{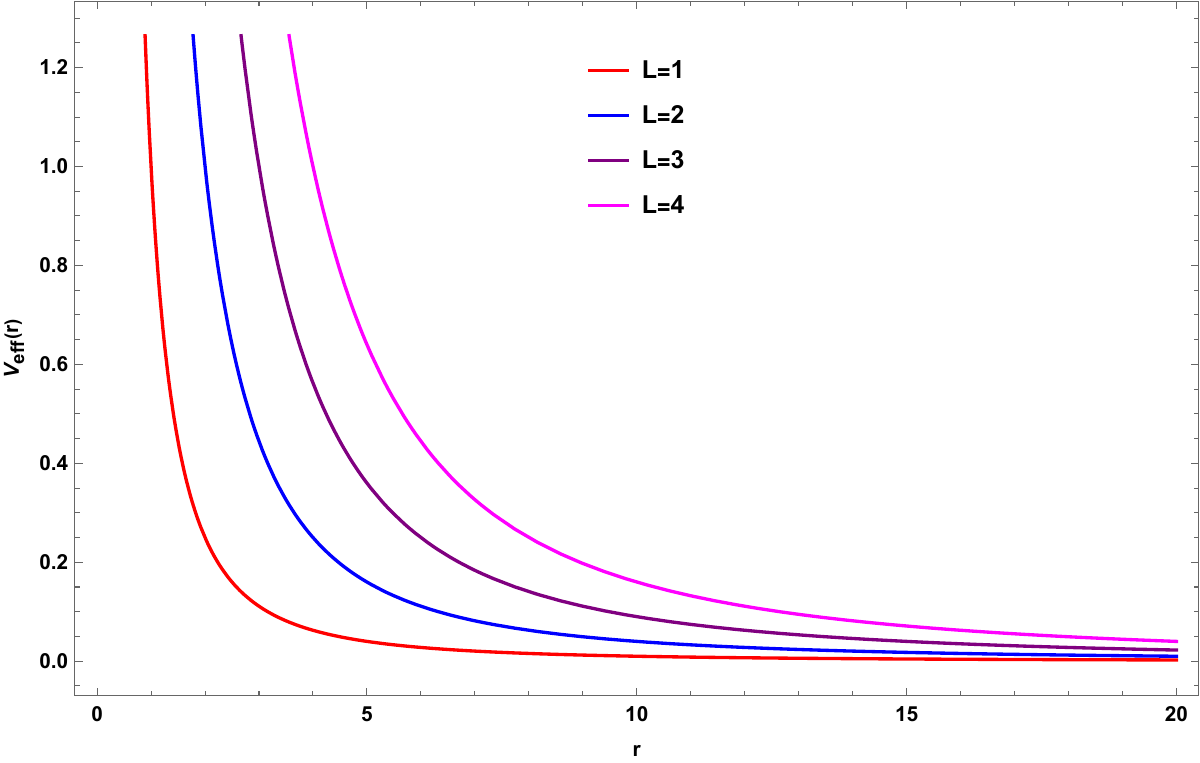}}\\
\subfloat[]{\centering{}\includegraphics[width=0.7\linewidth]{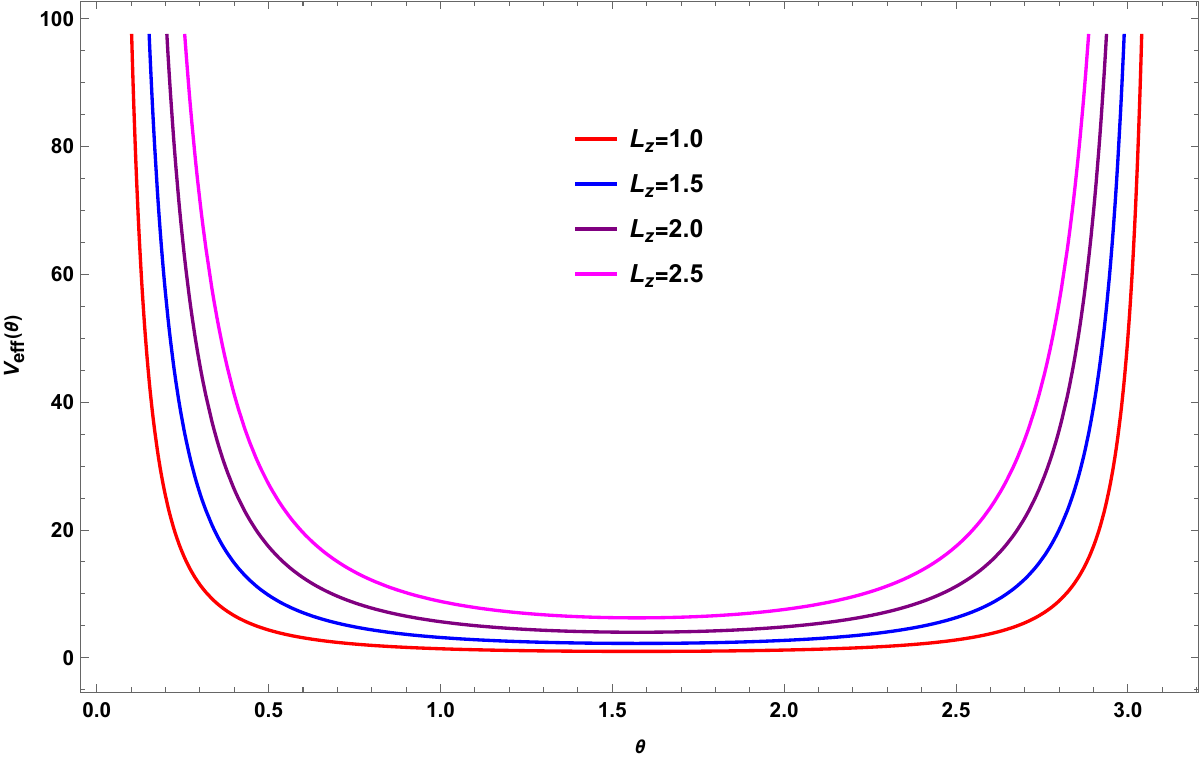}}\\
\caption{Illustration of the effective potential $V_\text{eff}(r)$ as a function of $r$ and $V_\text{eff}(\theta)$ as a function of $\theta$ for different values of $\mathrm{L}$ and $\mathrm{L}_z$.}
\label{fig:1}
\end{centering}
\end{figure}
\par\end{center}

\begin{center}
\begin{figure}[ht!]
\begin{centering}
\subfloat[$\alpha=0.9$]{\centering{}\includegraphics[width=0.7\linewidth]{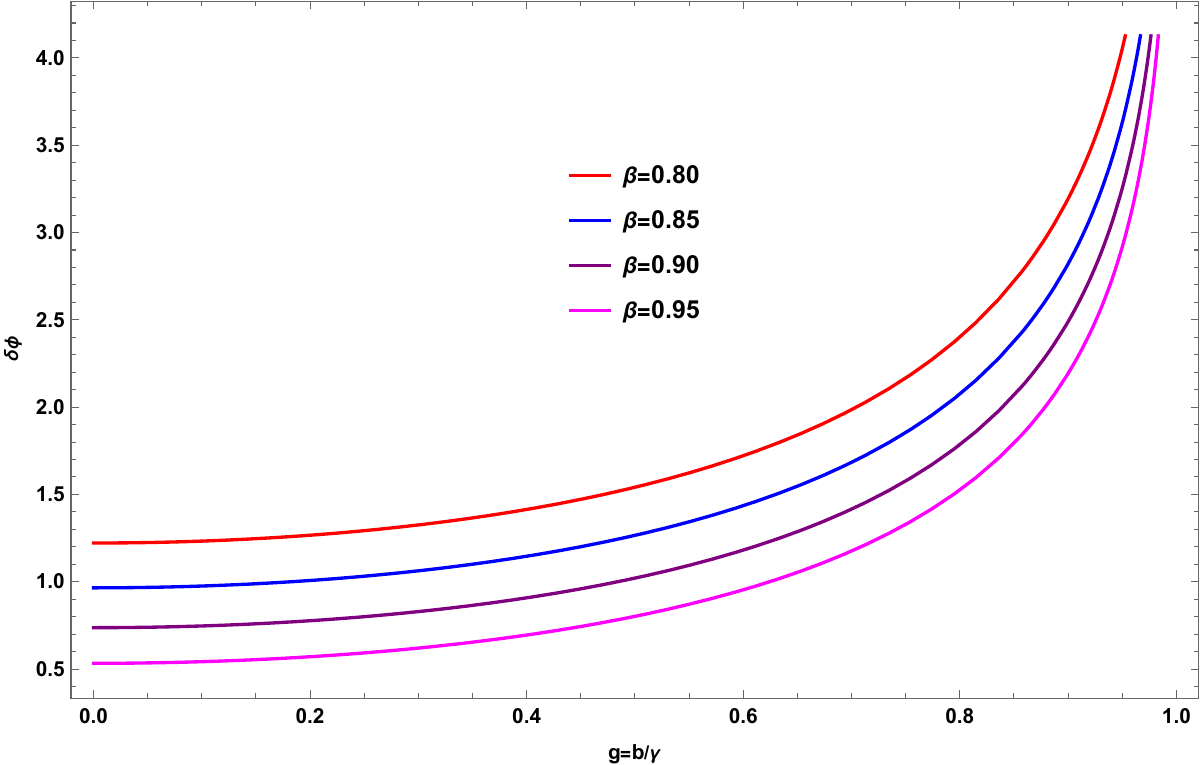}}\\
\subfloat[$\beta=0.95$]{\centering{}\includegraphics[width=0.7\linewidth]{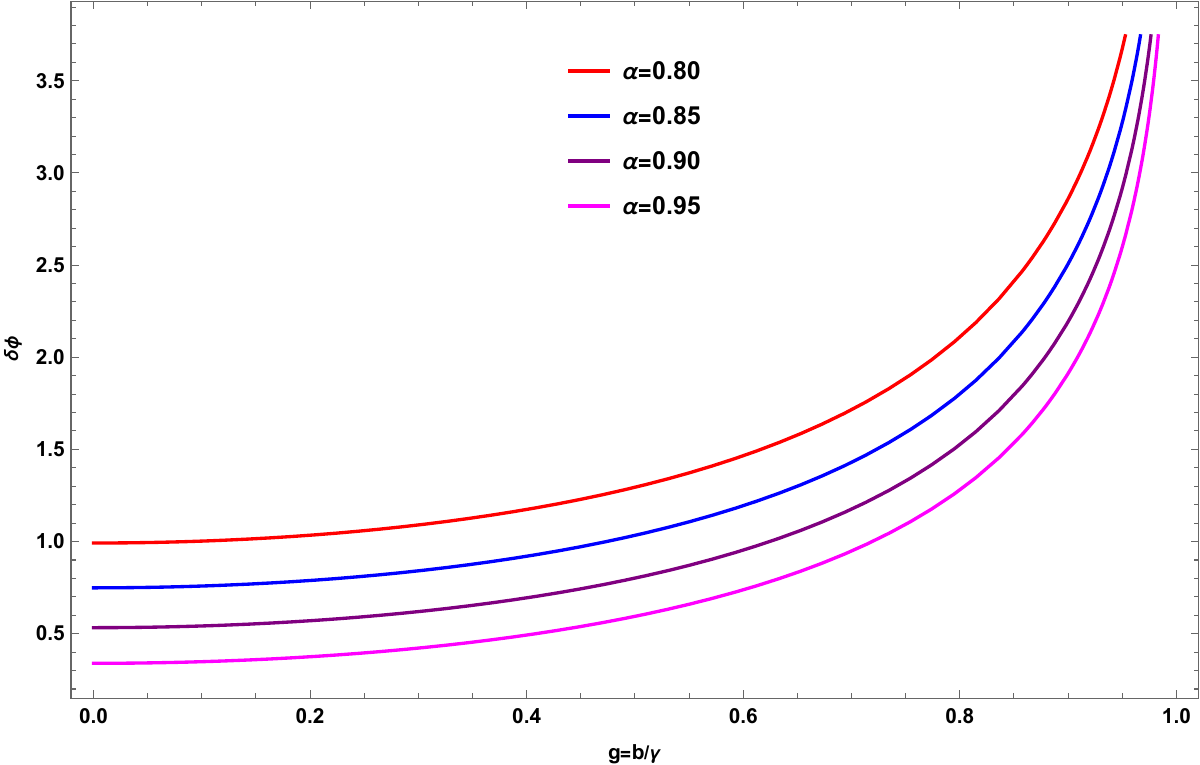}}\\
\subfloat[]{\centering{}\includegraphics[width=0.7\linewidth]{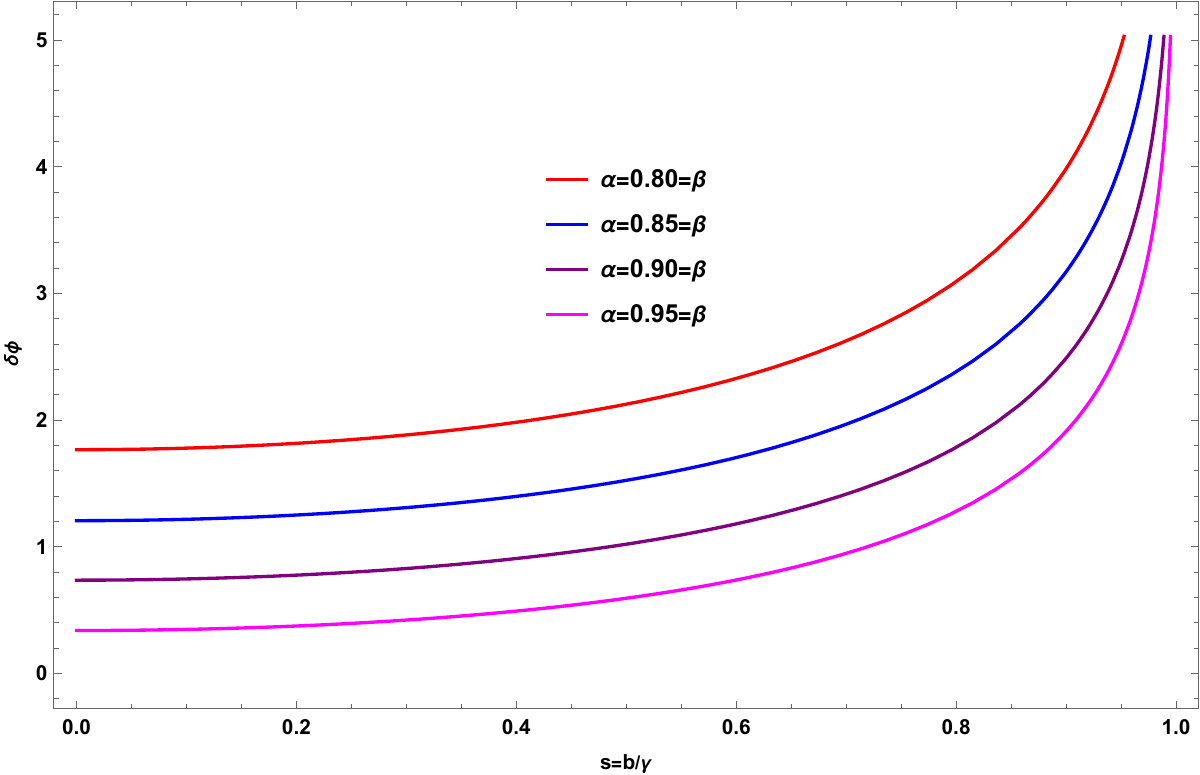}}
\caption{Illustration of the deflection angle $\delta\phi_{\text{Metric-I}}(\alpha,\beta,g)$ as a function of $g=b/\gamma$ for different values of $\alpha$ and $\beta$.}
\label{fig:2}
\end{centering}
\end{figure}
\par\end{center}

\begin{figure}[ht!]
    \centering
    \includegraphics[width=0.7\linewidth]{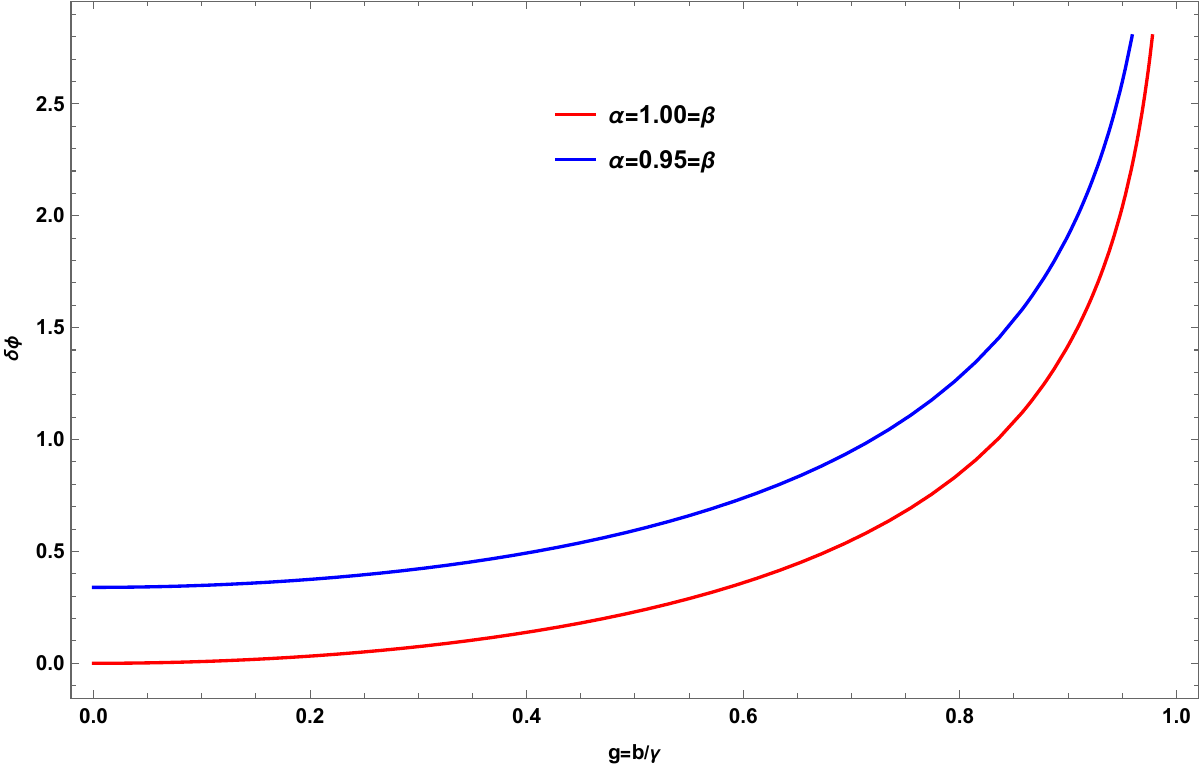}
    \caption{A comparison of the deflection angle $\delta\phi_{\text{Metric-I}}(\alpha,\beta,g)$. Colors $\to$ red: $\alpha=1=\beta$, blue: $\alpha=0.95=\beta$.}
    \label{fig:3}
\end{figure}

In Figure \ref{fig:1}, we present the effective potential $V_\text{eff}(r)$ for null geodesics ($\varepsilon=0$) given in Eq. (\ref{b6bb}) as a function of $r$ and $V_\text{eff}(\theta)$ given in Eq. (\ref{b6dd}) as a function of $\theta$ for different values of $\mathrm{L}$ and $\mathrm{L}_z$, illustrating how these parameters influence the shape of these effective potentials.

In Figure \ref{fig:2}, we present the deflection angle of photon light for various values of the GM parameter $\alpha$ and the CS parameter $\beta$. The graph clearly illustrates that as the value of the CS parameter $\beta<1$ increases; the deflection angle diminishes. A similar trend is observed when the value of the GM parameter $0 < \alpha<1$ increases. Furthermore, when both parameters, $0< \alpha \leq 1$ and $0 < \beta \leq 1$, increase simultaneously, the deflection angle also diminishes.

Additionally, Figure \ref{fig:3} presents a comparison of the deflection angle for different scenarios of the wormhole geometry. This comparison illustrates how the deflection angle varies under different conditions, such as the presence or absence of the GM and CS. Notably, we observe that the deflection angle is larger in scenarios where both the GM and CS are present, compared to cases where these topological parameters are absent. This behavior suggests that the combined effects of the GM, characterized by the parameter $\alpha$, and the CS, characterized by the parameter $\beta$, enhance the deflection of light in the gravitational field of the selected wormhole space-time. The increase in the deflection angle can be attributed to the collective influence of these topological features, which modify the curvature of space-time in the region surrounding the wormhole, leading to a stronger bending of light.

\section{Discussion} \label{sec3}

Now, we compare our results Eqs. (\ref{b19}) or (\ref{f10}) with a few known results obtained in the literature and show how the global monopole and the CS parameter together influence the deflection angle of photon ray in the gravitational field produced by the selected wormhole space-times. 

In Ref. \cite{KN}, the authors obtained the deflection angle of light photon in an Ellis wormhole geometry. In terms of impact parameter $\gamma$ and throat radius $b$ (using our notations), this angle is given by
\begin{eqnarray}
\delta\phi=\frac{\pi}{4}\,g^2+\frac{9\,\pi}{64}\,g^4+\mathcal{O} (g)^6.\label{c1}
\end{eqnarray}

In Ref. \cite{bb22}, the authors presented the deflection angle of photon light in a topologically charged Ellis-Bronnikov-type wormhole geometry.  In terms of impact parameter $\gamma$ and throat radius $b$ (using our notations), this angle is given by
\begin{eqnarray}
\delta\phi=\frac{\alpha^2\,\pi}{2}+\frac{\pi}{4}\,g^2+\frac{9\,\pi}{64}\,g^4+\frac{\pi\,\alpha^2}{8}\,g^2
+\mathcal{O} (g)^6.\quad\quad\label{c3}
\end{eqnarray}

In Ref. \cite{FA}, the deflection angle of photon light in Morris-Thorne-type wormhole, featuring only a CS effect was obtained. In terms of the impact parameter $\gamma$ and throat radius $b$ (using our notations), this deflection angle is given by
\begin{eqnarray}
\delta\phi=\left(\frac{1}{\beta}-1\right)\,\pi+\frac{\pi\,\beta}{4}\,g^2+\frac{9\,\pi\,\beta^3}{64}\,g^4
+\beta^5\,\mathcal{O} (g)^6.\label{c4}
\end{eqnarray}

In the limit when $\alpha=1$ and $\beta=1$, effectively removing the influence of both the GM and the CS parameters, the metrics in Eqs. (\ref{b1}) and (\ref{f1}) reduces to EBMT wormhole space-time. Consequently, the deflection angle of photon light, as derived in the current study in Eqs. (\ref{b19}) and/or (\ref{f6}), simplifies to the same form given by:
\begin{eqnarray}
\delta\phi=\frac{\pi}{4}\,g^2+\frac{9\,\pi}{64}\,g^4+\mathcal{O} (g)^6\label{c7}
\end{eqnarray}
which is similar to the expression in Eq. (\ref{c1}) obtained in Ref. \cite{KN}.

By comparing the result for the deflection angle presented in Eq. (\ref{b19}) with the one given in Eq. (\ref{c1}) from Ref. \cite{KN}, we derive the following relationship between the deflection angles for the EBMT-type wormhole with and without topological defects:
\begin{equation}
    \delta\phi_\text{Metric-I or II}=\left(\frac{1}{\alpha\,\beta}-1\right)\,\pi+\frac{1}{\alpha\,\beta}\,\delta\phi_\text{EBMT}.\label{comparison}
\end{equation}

Moreover, in the limit when $\alpha=1$, effectively removing the influence of the GM from the metrics in Eqs. (\ref{b1}) and (\ref{f1}), the deflection angle of photon light, as derived in the current study in Eqs. (\ref{b19}) and/or (\ref{f10}), simplifies to the same form given by:
\begin{eqnarray}
\delta\phi=\left(\frac{1}{\beta}-1\right)\,\pi+\frac{\pi}{4\,\beta}\,g^2+\frac{9\,\pi}{64\,\beta}\,g^4
+\frac{1}{\beta}\,\mathcal{O} (g)^6.\label{c6}
\end{eqnarray}
which is different from the expression in Eq. (\ref{c4}) obtained in Ref. \cite{FA}. 

Finally, in the limit when $\beta = 1$, effectively eliminating the influence of the CS parameter, the wormhole metrics in Eqs. (\ref{b1}) and (\ref{f1}) reduces to topologically charged EBMT-type wormhole space-time of different forms \cite{JRN,bb22,JG}. Consequently, the deflection angle of photon light can be obtained from Eqs. (\ref{b19}) and (\ref{f10}) by setting $\beta = 1$, yielding the following same expression ($g=b/\gamma$):
\begin{eqnarray}
\delta\phi_{\text{Metric-I or II}}=\left(\frac{1}{\alpha}-1\right)\,\pi+\frac{\pi}{4\,\alpha}\,g^2+\frac{9\,\pi}{64\,\alpha}\,g^4+\frac{1}{\alpha}\,\mathcal{O}(g)^6.\quad\quad\label{c5}
\end{eqnarray}
The results obtained above are fundamentally different from the expression in Eq. (\ref{c3}) obtained in Ref. \cite{bb22}.

\begin{center}
    \large{\bf Special Case Corresponds to $b=0$.}
\end{center}

A special case where $b=0$ is considered in the metrics provided in Eqs. (\ref{b1}) and (\ref{f1}) within this discussion. In the limit where $b = 0$, the metrics (\ref{b1}) and (\ref{f1}) simplifies to a point-like GM embedded with a CS \cite{MBAV,FA,AV,Vilekan,BL}. The resulting space-time configuration is no more a wormhole but a metric far from the core of GM that possesses a non-trivial topological feature including CS. This reduction yields the following form for the metrics as,
\begin{eqnarray}
   ds^2_{\text{GM+CS-I}}&=&-dt^2+\frac{dr^2}{\alpha^2}+r^2\,(d\theta^2+\beta^2\,\sin^2 \theta\,d\phi^2),\label{d1a}\\
   ds^2_{\text{GM+CS-II}}&=&-dt^2+dr^2+\alpha^2\,r^2\,(d\theta^2+\beta^2\,\sin^2 \theta\,d\phi^2).\label{d1b}
\end{eqnarray}

Following the previous procedure, the geodesics equations are given by 
\begin{eqnarray}
    &&\text{GM+CS-I:}\nonumber\\
    &&\left(\frac{dt}{d\tau}\right)^2=\mathrm{E}^2,\quad\quad\quad\quad\quad\quad \left(\frac{dr}{d\tau}\right)^2=\alpha^2\,\left(\mathrm{E}^2-\frac{\mathrm{L}^2}{r^2}\right),\nonumber\\
    &&\left(\frac{d\phi}{d\tau}\right)^2=\frac{\mathrm{L}^2_z}{\beta^2\,r^4\,\sin^2 \theta},\quad\quad 
    \left(\frac{d\theta}{d\tau}\right)^2=\frac{\mathrm{L}^2}{r^4}-\frac{\mathrm{L}^2_z}{r^4\,\sin^2 \theta},\label{d4aa}\\
    &&\text{GM+CS-II:}\nonumber\\
    &&\left(\frac{dt}{d\tau}\right)^2=\mathrm{E}^2,\quad\quad\quad\quad\quad\quad \left(\frac{dr}{d\tau}\right)^2=\left(\mathrm{E}^2-\frac{\mathrm{L}^2}{r^2}\right),\nonumber\\
    &&\left(\frac{d\phi}{d\tau}\right)^2=\frac{\mathrm{L}^2_z}{\alpha^2\,\beta^2\,r^4\,\sin^2 \theta},\quad
    \left(\frac{d\theta}{d\tau}\right)^2=\frac{\mathrm{L}^2}{\alpha^2\,r^4}-\frac{\mathrm{L}^2_z}{\alpha^2\,r^4\,\sin^2 \theta}.\quad\label{d4bb}
\end{eqnarray}

Employing the similar argument and the procedure done earlier, one can find the angular deflection of photon ray as $\delta\phi=\Delta\phi-\pi$ where $\Delta\phi$ is now given by
\begin{eqnarray}
    &&\Delta\phi_{\text{GM+CS-I}}=\frac{2\,\gamma}{\alpha\,\beta}\,\int^{\infty}_{r_0=\gamma}\,\frac{dr}{r\,\sqrt{r^2-\gamma^2}}=\frac{\pi}{\alpha\,\beta}.\label{d2a}\\
   &&\Delta\phi_{\text{GM+CS-II}}=\frac{2\,\gamma}{\alpha\,\beta}\,\int^{\infty}_{r_0=\gamma}\,\frac{dr}{r\,\sqrt{r^2-\gamma^2}}=\frac{\pi}{\alpha\,\beta}.\label{d2b}
\end{eqnarray}

Thereby, using the result (\ref{d2a}) or (\ref{d2b}), we find the angular deflection given as follows:
\begin{equation}
    \delta\phi_\text{GM+CS}=\left(\frac{1}{\alpha\,\beta}-1\right)\,\pi.\label{d3}
\end{equation}

From the above expression, it is clear that the combined presence of a GM charge and CS within the space-time configurations leads to a deviation in the photon light trajectories. This alteration in the photon path results in a measurable deflection angle, which is influenced by these topological defects. An interesting observation is that when the wormhole throat radius $b \neq 0$, the deflection angle derived from Metric-I or Metric-II is greater than the result for conical space-times (\ref{d1a}) and/or (\ref{d1b}), that is, $\delta\phi_{\text{Metric-I or II}}>\delta\phi_\text{GM+CS}$.

Therefore, we can express the deflection angle given in Eq. (\ref{comparison}) using Eq. (\ref{d3}) as follows:
\begin{equation}
    \delta\phi_\text{Metric-I or II}=\delta\phi_\text{GM+CS}+\frac{1}{\alpha\,\beta}\,\delta\phi_\text{EBMT}.\label{comparison2}
\end{equation}
The total deflection angle (\ref{comparison2}) is the result of two contributions. The first originates from the gravitational field of a non-wormhole curved space-time, which includes the effects of both the GM and the CS. The second contribution, scaled by a factor of $\frac{1}{\alpha\,\beta}$, represents the deflection angle caused by the EBMT wormhole metric. Together, these two components determine the overall deflection angle. 
} 

\begin{figure}[ht!]
    \centering
    \includegraphics[scale=0.32]{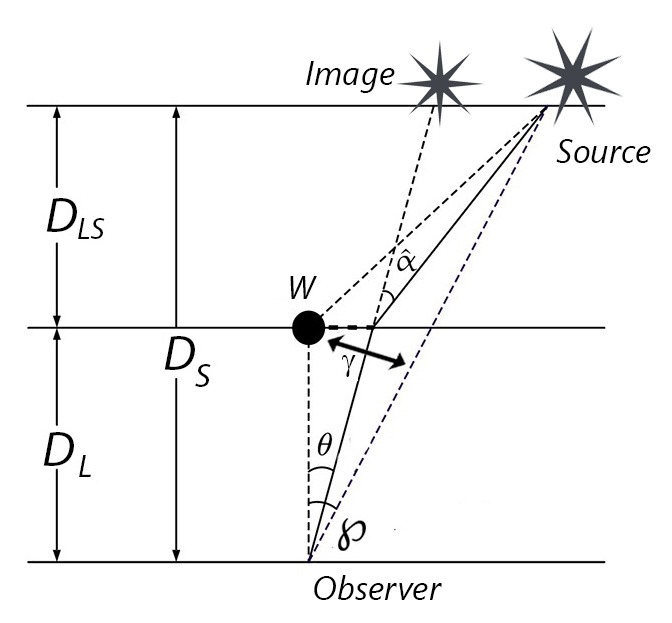}
    \caption{The lensing geometry. Here $W$ corresponds to a massless WH as a lens, $\alpha$ is the deflection angle, $\phi$ is the angle between the wormhole and the light source, $\gamma$ is the impact parameter that is perpendicular to the dotted line connecting the source and the observer. $D_L$, $D_S$, and $D_{LS} = D_S - D_L$ are the angular diameter distances. To enhance conceptual clarity, we assume $D_{LS} \approx D_L$ and $D_{LS}/D_S \approx 1/2$. Under a small angle approximation, where $\sin \theta \approx \theta$ for $\theta \ll 1$, one can obtain $\gamma/D_L \approx \theta$.}
        \label{fig:5}
\end{figure}

\section{Observational Consequences}
\label{sec_observational}

While our theoretical analysis of the deflection angle provides fundamental insights into Ellis-Bronnikov-Morris-Thorne wormholes with GM and CS, it is essential to translate these results into observable astrophysical signatures. In this section, we analyze how the distinctive features of our model would manifest in observable quantities, comparing them with standard black hole lensing. A typical gravitational lensing WH geometry is illustrated in Figure \ref{fig:5}.

Starting from the deflection angle derived in our work, we can establish the lens equation that relates source position $\wp$ to image position $\theta$:

\begin{equation}
   \wp=\theta-\frac{D_{LS}}{D_S}\,\hat{\alpha},\quad\quad D_{LS}=D_S-D_L,\label{ss1}
\end{equation}

where $\hat{\alpha}$ is the deflection angle. Using our result from Eq. (\ref{comparison2}), and considering first-order term, we have:

\begin{eqnarray}
    &&\delta\phi_\text{Metric-I or II}=\hat{\alpha},\quad \delta\phi_\text{GM+CS}=\hat{\alpha}_0,\quad \delta\phi_\text{EBMT}=\frac{\pi\,b^2}{4\,\gamma^2},\nonumber\\
    &&\hat{\alpha}=\hat{\alpha}_0+\frac{\pi\,b^2}{4\,\alpha\,\beta\,\gamma^2},\quad\quad \hat{\alpha}_0 = \frac{(1-\alpha\,\beta)}{\alpha\,\beta}\,\pi,\label{ss2}
\end{eqnarray}

where $\hat{\alpha}_0$ represents the contribution from the topological defects (GM plus CS) alone, and the second term represents the wormhole throat contribution. Using the approximation $\gamma \approx \theta\,D_L$, where $\gamma$ is the impact parameter for photon ray, the lens equation becomes:

\begin{equation}
   \wp=\theta-\frac{D_{LS}}{D_S}\,\hat{\alpha}_0-\frac{D_{LS}}{D_S}\,\frac{\pi\,b^2}{4\,\alpha\,\beta\,\theta^2\,D^2_L}.\label{ss3}
\end{equation}

At this stage, it is essential to note that the Einstein ring typically forms when the lens, source, and observer are perfectly aligned. In such a configuration, the lensing source is considered as a particle-like, mass object. However, in our case, we have considered a massless point source, where no Newtonian potential exists. This contrasts with traditional lensing cases, as discussed in Refs. \cite{MS,LHL}, where a point object with positive or negative mass generates a non-zero Newtonian potential, in the weak field approximation. The absence of such a potential in our massless point source leads to a fundamentally different lensing behavior. While traditional lensing relies on the gravitational field of the lensing object to bend light, our massless source’s lensing effect is shaped by the geometry of the wormhole itself, not by a conventional gravitational potential. This introduces a novel deflection pattern, departing from classical gravitational lensing predictions.

As shown in Figure \ref{fig:5}, one can approximately express $ \theta \approx \gamma/D_L$, where we assume $\sin \theta \approx \theta$ for small angle, $\theta \ll 1$. This approximation holds because the image of light is expected to be very close to the source. This subtle distinction, which arises from the massless nature of point source, significantly alters the lensing dynamics, marking a departure from the traditional lensing behavior governed by point mass (either positive or negative) source.

The above equation (\ref{ss3}) can be rewritten to the following cubic equation form:
\begin{equation}
   \theta^3-\chi\,\theta^2-\frac{\varsigma}{\alpha\,\beta}=0,\label{ss4}
\end{equation}

where we set
\begin{equation}
   \chi=\wp+\frac{D_{LS}}{D_S}\,\hat{\alpha}_0,\quad\quad \varsigma=\frac{\pi\,b^2}{4\,D^2_L}\,\frac{D_{LS}}{D_S}.\label{ss5}
\end{equation}

The discriminant that determines the number of real solutions (corresponding to images) is (see appendix A in Ref. \cite{LHL}):
\begin{equation}
   \Delta=\frac{81\,\varsigma^2}{\alpha^2\,\beta^2}+\frac{4\,\chi^3\,\varsigma}{\alpha\,\beta}>0.\label{discriminant}
\end{equation}

Since $\Delta > 0$ for our parameter range, there is always one real solution to Eq. (\ref{ss4}), meaning that our wormhole model produces a single image of the source. This contrasts sharply with black hole lensing, which always produces two images when the source, lens, and observer are not perfectly aligned. On the other hand, the magnification of a gravitational lens is defined as \cite{MS,LHL}:

\begin{equation}
   \mu = \left|\frac{\wp}{\theta}\,\frac{d\wp}{d\theta}\right|^{-1}.\label{ss6}
\end{equation}

From the lens equation (\ref{ss3}), the total magnification is given by

\begin{equation}
   \mu_\text{tot} =\left|\left(1-\frac{D_{LS}}{D_S}\,\hat{\alpha}_0-\frac{\varsigma}{\alpha\,\beta\,\theta^3}\right)\left(1 + \frac{2\,\varsigma}{\alpha\,\beta\,\theta^3}\right)\right|^{-1},\label{ss7}
\end{equation}
where $\varsigma$ is defined in Eq. (\ref{ss5}). On the other hand, the tangential and radial magnifications are given by \cite{MS,LHL}
\begin{align}
\mu_{\tan }&=\left|\frac{\wp}{\theta}\right|^{-1}=\left|1-\frac{\pi(1-\alpha\, \beta)\,D_{LS}}{D_S\, \alpha\, \beta\, \theta}-\frac{1}{4} \frac{\pi\, b^2\,D_{LS}}{\alpha\, \beta\, D_L^2\, D_S\, \theta^3}\right|^{-1},    
\end{align}
and
\begin{align}
\mu_{\mathrm{rad}}=\left|\frac{d \wp}{d \theta}\right|^{-1}=\left|1+\frac{1}{2} \frac{\pi\, b^2\,D_{LS}}{\alpha\, \beta\, D_L^2\, D_S\, \theta^3}\right|^{-1}.
\end{align}

Figure~\ref{fig:magnification} illustrates the differential magnification patterns produced by the EBMT wormhole configuration with topological defects. Contrary to Schwarzschild black hole systems \cite{MS}, which exhibit characteristic double-peak patterns with divergent magnification near alignment, the wormhole model demonstrates a fundamentally different signature: a symmetric U-shaped profile with minimal central magnification. Several quantitative observations warrant attention: (1) The total magnification $\mu_{\mathrm{tot}}$ remains remarkably small ($\sim 10^{-38}$) across the central angular range, increasing modestly only at large angular displacements; (2) Both tangential and radial components ($\mu_{\mathrm{tan}}$ and $\mu_{\mathrm{rad}}$) demonstrate complementary behavior that effectively counterbalances to produce the characteristic flat central region; (3) The total magnification profile maintains consistent morphology regardless of the topological parameters $\alpha$ and $\beta$. This magnification structure represents a distinctive observational discriminant between classical black hole lensing and wormhole configurations with topological defects. The negligible central magnification implies that light paths through such wormhole systems would produce significantly dimmer lensed images compared to black hole counterparts, potentially providing empirical constraints on the existence of such exotic space-time structures through precision gravitational lensing surveys.

In the limit where $\alpha=1$ and $\beta=1$, we recover EBMT wormhole metric without topological defects. In that case, from Eq. (\ref{ss2}) we find $\hat{\alpha}_0=0$, and hence, the lens equation from Eq. (\ref{ss4}) using Eq. (\ref{ss5}) becomes
\begin{equation}
    \theta^3-\wp\,\theta^2-\varsigma=0.\label{lens}
\end{equation}
And that the magnification from Eq. (\ref{ss7}) becomes
\begin{equation}\label{ss9}
   \mu_\text{EBMT}=\left|(1-z)(1+2\,z)\right|^{-1}=\left|1+z-2\,z^2\right|^{-1},
\end{equation}
where $z=\frac{\varsigma}{\theta^3}$.

A comparison of the current magnification Eq. (\ref{ss7}) with that for an EBMT wormhole result (\ref{ss9}) is as follows:
\begin{equation}
    \frac{\mu_\text{tot}}{\mu_\text{EBMT}} =\frac{\left|1+z-2\,z^2\right|}{\left|\left(1-\frac{D_{LS}}{D_S}\,\hat{\alpha}_0-\frac{z}{\alpha\,\beta}\right)\,\left(1 + \frac{2\,z}{\alpha\,\beta}\right)\right|}< 1.\label{comparison3}
\end{equation}

The expression above highlights that the inclusion of GM, characterized by $\alpha$, and CS, represented by $\beta$, in the EBMT wormhole geometry significantly influences the magnification of images. This alteration suggests that the properties of GM and CS play a crucial role in modifying the gravitational lensing effects, thereby changing how light is focused or distorted in such a space-time structure. Consequently, the magnification and the overall image formation are directly impacted by the parameters $\alpha$ and $\beta$, revealing the complex interplay between matter, curvature, and light propagation in this context.  

\section{Summary and conclusions}\label{sec4}

In this study, we conducted a theoretical investigation of the gravitational lensing effects exhibited by an EBMT-type traversable wormhole in the presence of a GM and CM. We focused on analyzing how these topological defects influence the deflection of light passing through the wormhole. To achieve this, we first derived the equations of motion for photons in the combined gravitational field of the wormhole, GM, and CS. We then employed these equations to determine the trajectories of light rays and calculate the deflection angle, taking into account the parameters of the wormhole (throat radius $b$), the GM (parameter $\alpha$), and the CS (parameter $\beta$).

Our analysis revealed that the combined effects of the GM and CS significantly alter the deflection angle of photon light compared to cases where these topological defects are absent or considered individually. This modification is evident in the expression for the deflection angle, given by Eq. (\ref{b19}). This equation demonstrates the explicit dependence of the deflection angle on both the GM parameter $\alpha$ and the CS parameter $\beta$. Furthermore, we presented a graphical analysis of the deflection angle in Figures \ref{fig:2} and \ref{fig:3}. Figure \ref{fig:2} illustrates how the deflection angle varies with the parameters $\alpha$ and $\beta$. We observed that as the values of $\alpha < 1$ and $\beta < 1$ increase, the deflection angle decreases. This indicates that the presence of the GM and CS, within the specified parameter range, tends to reduce the overall bending of light. Figure \ref{fig:3} provides a comparison of the deflection angle for two scenarios: an EBMT without any topological defects, and an EBMT with both GM and a CS. This comparison clearly shows that the deflection angle is largest when both the GM and CS are present. 

The deflection angles obtained in Eqs. \eqref{b19} and \eqref{f10} showed same contributions from the combined effects of the GM ($\alpha$), CS ($\beta$), wormhole throat radius ($b$), and impact parameter ($\gamma$). The first term, $\left(\frac{1}{\alpha \beta}-1\right) \pi$ in Eq. (\ref{comparison2}), arose solely due to the presence of the GM and CS, independent of the wormhole geometry. This term represented the fundamental shift in the deflection angle induced by these topological defects. The second term, $\frac{\pi}{4\,\alpha\,\beta}\,g^2$ emerged from the combined influence of the GM, CS, wormhole throat radius ($b$), and impact parameter ($\gamma$). This correction reflected how the presence of the wormhole altered light bending in addition to the effects of the topological defects. The third term, $\frac{9\,\pi}{64\,\alpha\,\beta}\,g^4$ arose from higher-order gravitational effects and integrated nonlinear modifications due to the interplay of the wormhole throat, GM, and CS. Finally, the last term, involving $\mathcal{O}(g)^6$, represented further corrections that became relevant in the weak-field approximation when considering higher-order gravitational lensing effects. 

\begin{figure}[ht!]
   	\centering
   	\quad\quad\quad\includegraphics[scale=0.65]{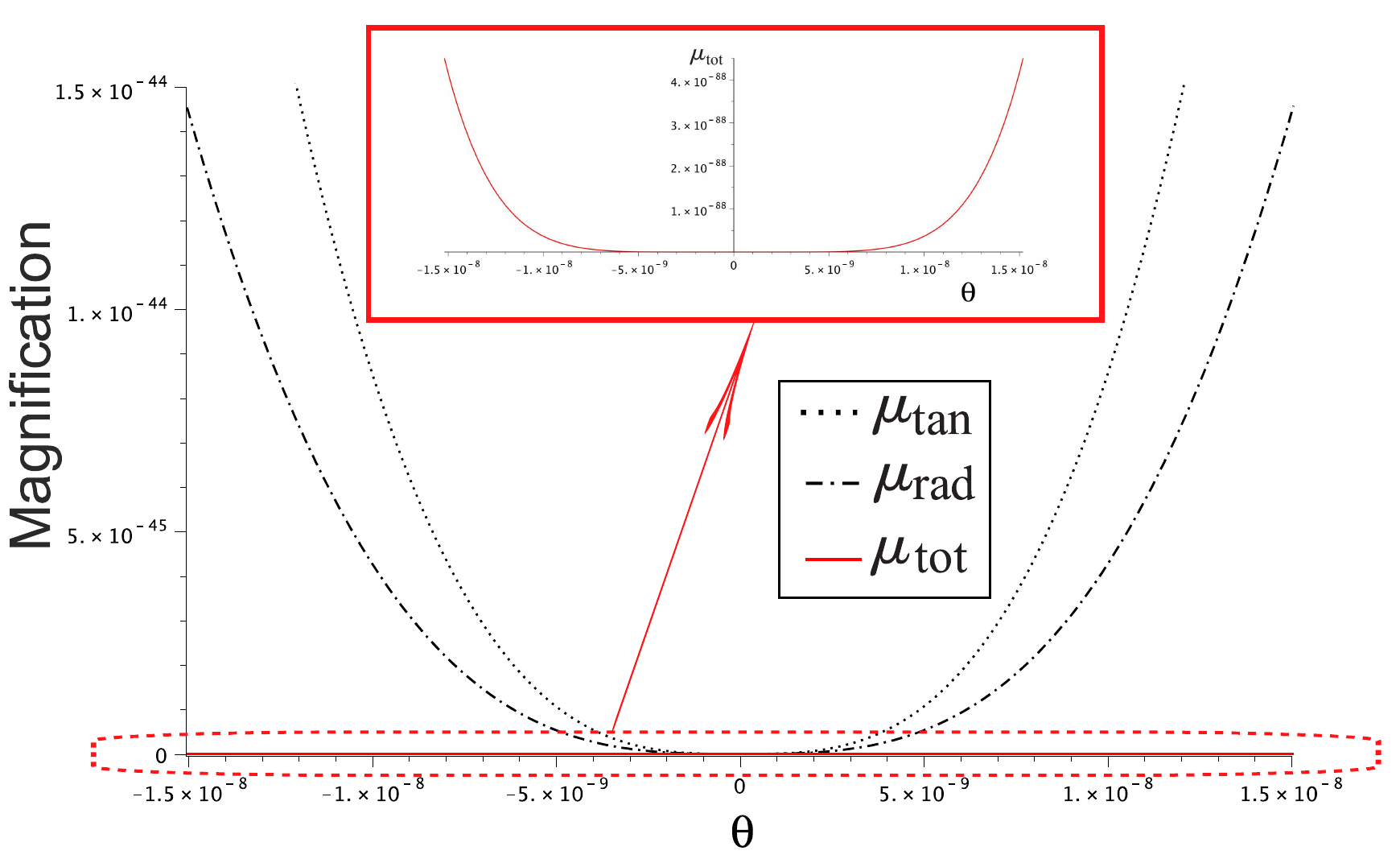}
   	\captionsetup{width=\textwidth}
   	\caption{Magnification properties of EBMT wormhole with GM and CS topological defects. The figure presents tangential $\mu_{\mathrm{tan}}$ (dotted curves), radial $\mu_{\mathrm{rad}}$ (dash-dotted curves), and total $\mu_{\mathrm{tot}}$ (solid line) magnifications as functions of angular image position $\theta$. The computational parameters are fixed at $\alpha=\beta=0.95$, $b=4 \times 10^{30}m$, $D_{\mathrm{s}}=0.05 \mathrm{Mpc}$, and $D_{\mathrm{L}}=0.01 \mathrm{Mpc}$. Angular measurements span from $-1.5\times10^{-8}$ to $+1.5\times10^{-8}$ radians, corresponding to $-3.14 \times 10^{-3}$ to $+3.14 \times 10^{-3}$ arcseconds. The inset (red box) provides detailed visualization of the total magnification function across the full angular range, highlighting the characteristic U-shaped profile distinctive of wormhole lensing phenomena.}
   	\label{fig:magnification}
   \end{figure}

Our results highlight the significant role that GM and CS play in modifying the gravitational lensing properties of EBMT-type wormholes. Specifically, the analysis of the deflection angle, as shown in Fig. \ref{fig:2}, shows that varying either the CS parameter while keeping GM fixed, or varying GM while keeping CS fixed, both impact the deflection angle. This effect becomes even more pronounced when both GM and CS are varied simultaneously. The comparative plot in Fig. \ref{fig:3} offers a clear visualization of these combined effects, illustrating the deflection angle for the EBMT wormhole with and without GM and CS, showing a substantial enhancement when both are present. This finding suggests that GM and CS together exert a more significant gravitational influence on light bending than when these topological parameters are absent, making them a crucial factor in determining the observational signatures of wormhole space-times. Moreover, our analysis of the gravitational lensing characteristics of EBMT wormholes with topological defects has also illuminated distinctive magnification signatures, as demonstrated in Fig.~\ref{fig:magnification}. The remarkably suppressed total magnification ($\sim 10^{-38}$) exhibited by these exotic spacetime structures contrasted sharply with the divergent behavior typically associated with Schwarzschild black hole lensing systems. Furthermore, the characteristic U-shaped magnification profile maintained its morphological integrity irrespective of variations in the topological parameters $\alpha$ and $\beta$, suggesting a remarkable observational discriminant for future astrophysical surveys. 

Building upon the foundation laid by this investigation, several promising future research emerge. One intriguing direction is to study the strong field limit of gravitational lensing of EBMT with topological defects. While our study focused on the weak field approximation, where light rays do not pass close to the wormhole throat, exploring the strong field regime could unveil more intricate lensing effects, potentially offering stronger observational signatures for these objects \cite{isz85,RS}. This could involve analyzing the behavior of light rays that undergo multiple orbits around the wormhole before reaching the observer, leading to the formation of relativistic images and potentially revealing unique features associated with the combined effects of the wormhole and topological defects. Another compelling project is to extend this analysis to different wormhole geometries. Our study specifically considered the static Ellis wormhole, but various other wormhole models exist, such as rotating wormholes \cite{isz18} or wormholes with dynamic geometries \cite{isz66}. Investigating how the presence of GM and CM affects light deflection in these alternative space-times could provide a more comprehensive understanding of the interplay between wormhole geometry and topological defects in shaping gravitational lensing phenomena. Finally, a crucial step in advancing this research is to analyze observational data from gravitational lensing surveys in light of our theoretical predictions. By comparing the predicted lensing patterns with actual observations, one can potentially identify signatures of wormholes and topological defects in astrophysical data \cite{isz21,isz88}. This could involve searching for characteristic distortions in the images of distant galaxies or seeking evidence of multiple images with specific angular separations and magnification ratios. Such future observational studies could provide crucial evidence for the existence of these exotic objects and offer valuable insights into the nature of gravity and the early universe.

{\small
\section*{Acknowledgments}

The authors express their appreciation to the editor for meticulously managing the manuscript, and to the anonymous referee whose insightful and constructive feedback significantly improved the quality and clarity of this study. F.A. acknowledges the Inter University Centre for Astronomy and Astrophysics (IUCAA), Pune, India for granting visiting associateship. \.{I}.~S. thanks T\"{U}B\.{I}TAK, ANKOS, and SCOAP3 for funding and acknowledges the networking support from COST Actions CA22113, CA21106, and CA23130.

\section*{Appendix A : Matter-Energy Distribution and The Energy Conditions}

\setcounter{equation}{0}
\renewcommand{\theequation}{A.\arabic{equation}}

For the Metric-I given by the line-element (\ref{b1}), the non-zero components of the Einstein tensor $G^{\mu}_{\,\nu}$ are given by
\begin{eqnarray}
    &&G^{t}_{t}=-\left(\frac{1-\alpha^2}{r^2}-\frac{\alpha^2\,b^2}{r^4}\right),\quad G^{r}_{r}=-\left(\frac{1-\alpha^2}{r^2}+\frac{\alpha^2\,b^2}{r^4}\right),\nonumber\\
    &&G^{\theta}_{\theta}=\frac{\alpha^2\,b^2}{r^4}=G^{\phi}_{\phi}.\label{A1}
\end{eqnarray}

Considering the energy-momentum tensor of the following form
\begin{equation}
    T^{\mu}_{\,\nu}=\mbox{diag}(-\rho,p_r,p_t,p_t),\label{A2}
\end{equation}
where $\rho$ is the energy-density, $p_r, p_t$  are the radial and tangential pressures, respectively.

From the Field Equations $G^{\mu}_{\,\nu}=T^{\mu}_{\,\nu}$, we find
\begin{eqnarray}
    \rho=\frac{1-\alpha^2}{r^2}-\frac{\alpha^2\,b^2}{r^4},\quad
    p_r=-\frac{1-\alpha^2}{r^2}-\frac{\alpha^2\,b^2}{r^4},\quad
    p_t=\frac{\alpha^2\,b^2}{r^4}.\quad\label{A3}
\end{eqnarray}

At the wormhole throat $r=b$, we find: 
\begin{eqnarray}
    &&\rho|_{r=b}=\frac{1-2\,\alpha^2}{b^2},\quad\quad
    \rho+p_r|_{r=b}=-\frac{2\,\alpha^2}{b^2}<0,\nonumber\\
    &&\rho+p_t|_{r=b}=\frac{1-\alpha^2}{b^2}>0,\quad\quad 
    \rho+p_r+2\,p_t=0.\label{A4}
\end{eqnarray}

Similarly, for the Metric-II given by the line-element (\ref{f1}), the non-zero components of the Einstein tensor $G^{\mu}_{\,\nu}$ are given by
\begin{eqnarray}
    &&G^{t}_{t}=-\frac{1}{\alpha^2\,r^2}+\frac{b^2}{r^4}+\frac{1}{r^2},\quad G^{r}_{r}=-\frac{1}{\alpha^2\,r^2}+\frac{1}{r^2}-\frac{b^2}{r^4},\nonumber\\
    &&G^{\theta}_{\theta}=\frac{b^2}{r^4}=G^{\phi}_{\phi}.\label{A5}
\end{eqnarray}

Thereby, using the energy-momentum tensor (\ref{A2}) and from the field equations, we find
\begin{eqnarray}
    \rho=\frac{1}{\alpha^2\,r^2}-\frac{b^2}{r^4}-\frac{1}{r^2},\quad p_r=-\frac{1}{\alpha^2\,r^2}+\frac{1}{r^2}-\frac{b^2}{r^4},\quad
    p_t=\frac{b^2}{r^4}.\quad\label{A6}
\end{eqnarray}
At the wormhole throat $r=b$, we find 
\begin{eqnarray}
    &&\rho|_{r=b}=\frac{1-2\,\alpha^2}{\alpha^2\,b^2},\quad
    \rho+p_r|_{r=b}=-\frac{2}{b^2}<0,\nonumber\\
    &&\rho+p_t|_{r=b}=\frac{1-\alpha^2}{\alpha^2\,b^2}>0,\quad
    \rho+p_r+2\,p_t=0.\label{A7}
\end{eqnarray}
From the relations (\ref{A4}) and (\ref{A7}), we observe that the energy density of the matter content remains positive for the range $0 <\alpha <1/\sqrt{2}$, but becomes negative when $1/\sqrt{2} < \alpha \leq 1$. Moreover, we see that the weak energy condition (WEC) is violated along the radial, while it satisfied along the tangential direction since $0 < \alpha < 1$.  

However, it is important to note that the chosen matter-energy content partially violates the null energy condition (NEC). For that we consider the following null vector field $k^{\mu}$ (where $k^{\mu}\,k_{\mu}=0$) for the space-times (\ref{b1}) and (\ref{f1}) as \cite{VAR1,VAR2}:
\begin{eqnarray}
    k^{\mu}&=&\left(1,\,\alpha\,\sqrt{1-b^2/r^2},\,0,\,0\right)\quad\quad (\mbox{Metric-I}),\label{A8}\\
    &=&\left(1,\,\sqrt{1-b^2/r^2},\,0,\,0\right)\quad\quad\quad (\mbox{Metric-II}).\label{A9}
\end{eqnarray}

For the chosen matter-energy content given in Eq. (\ref{A2}), we find the null energy condition as,
\begin{eqnarray}
    T^{\mu\nu}\,k_{\mu}\,k_{\nu}=\rho+p_r&=&-\frac{2\,\alpha^2\,b^2}{r^4}\quad\quad (\mbox{Metric-I}),\label{A10}\\
    &=&-\frac{2\,b^2}{r^4}\quad\quad\quad (\mbox{Metric-II}).\label{A11}
\end{eqnarray}
Hence, at the wormhole throat, $r=b$, we find $T^{\mu\nu}\,k_{\mu}\,k_{\nu}|_{r=b}<0$, thus, violates the NEC.

}

       \onecolumngrid
   
   \twocolumngrid

\end{document}